\documentclass[%
 reprint,
 amsmath,amssymb,
 aps,
]{revtex4-1}
\usepackage{mathrsfs}
\usepackage{graphicx}
\usepackage{dcolumn}
\usepackage{bm}

\begin{document}

\preprint{APS/123-QED}

\title{ Analysis of $^4$He($\gamma,p)$T and $^4$He($\gamma,n)$$^3$He Reactions
with Linearly Polarized Photons in the Energy Range up to 100\,MeV }

\author{Yu.P.Lyakhno}

\affiliation{%
 National Science Center "Kharkiv Institute of
Physics and Technology" \\ 61108, Kharkiv, Ukraine\\}%

\date{\today}

\begin{abstract}
In a number of investigations, one can find the data on the
$^4$He($\gamma$,p)T and $^4$He($\gamma$,n)$^3$He reaction cross
sections in the collinear geometry, which are due to spin {\it S}=1
transitions of the final-state particles. The ratio of the
differential cross section in the collinear geometry to the
differential reaction cross section at the nucleon emission angle
$\theta_N$=90$^\circ$, and specified by the {\it S}=0 electric
dipole transition at photon energies in the range 20$\le
E_{\gamma}\le$100\,MeV, is independent of the photon energy, within
the experimental error. In the meantime, experiments were made to
measure the asymmetry of the cross section $\Sigma(\theta_N)$, for
the mentioned reactions with linearly polarized photons. It has been
found that in the energy range between 20 and 90\,MeV, the
$\Sigma(\theta_N)$ value is also independent of the photon energy,
within the experimental error. These data are in agreement with the
assumption that transitions with spin {\it S}=1 can be due to the
contribution of $^3P_0$ states of the $^4$He nucleus, and are
inconsistent with the assumption that the spin-flip of the particle
system occurred during the reaction as a result of the meson
exchange current contribution. The available measured data on the
collinear geometry reaction cross sections and the ones on the
cross-section asymmetry of the reaction with linearly polarized
photons do not agree between themselves. The above mentioned
reactions seem to be more convenient for measuring the degree of
photon beams linear polarization than the deuteron
photodisintegration reactions.

\begin{description}

\item[PACS numbers]
21.30.-x; 21.45.-v; 24.70.+s.

\end{description}
\end{abstract}

\pacs{Valid PACS appear here}

\maketitle

\section{Introduction}

The total momentum and parity of the $^4$He nucleus J$^{\pi}$=0$^+$,
the two-body of the reactions, and also the absence of excited
states in the final nuclei make it possible to perform a detailed
multipole analysis in {\it E}1, {\it E}2 and {\it M}1 approximation.
In the course of the $^4$He($\gamma,p$)T and
$^4$He($\gamma,n$)$^3$He reactions the final-state spin of the
particle system can take two values: {\it S}=0 and {\it S}=1. The
main part total cross section of the  reaction  is contributed by
{\it S}=0 transitions. The contribution of these transitions is
explained by the direct nucleon knockout mechanism, the recoil
mechanism, and a number of exchange diagrams \cite{1}. In
Ref.\,\cite{2} the origination of {\it S}=1 transitions is explained
by the assumption that in the course of the reaction the spin-flip
of the hadronic particle system takes place, which results from the
contribution of meson exchange currents (MEC$^,$s). Based on the
realistic {\it NN} potential and 3{\it N} forces, Nogga et al.
\cite{3} have demonstrated that in the initial state the $^4$He
nucleus can also have the spin {\it S}=1 and {\it S}=2, and
possibly, the {\it S}=1 transitions may be due to the $^4$He nuclear
structure. Thus, the occurrence of multipole transitions may be of
different origin. The analysis of the available experimental data
along with obtaining new data is of considerable importance for
gaining new information on the nuclear structure and mechanisms of
nuclear reactions.

\section{ Multipole analysis of $^4$He($\gamma,p$)T and
$^4$He($\gamma,n$)$^3$He  reactions}

Complete expressions of expanding the differential cross sections
and cross-sectional asymmetry $\Sigma(\theta_N)$ in {\it E}1, {\it
E}2 and {\it M}1 multipoles for the $^4$He($\gamma$,p)T and
$^4$He($\gamma$,n)$^3$He reactions with linearly polarized photons
in the center-of-mass system can be written as:

\newcommand{\lambcross}{\lambda \hspace{-0.6em} {^-}}

\begin{eqnarray}\label{eq1} &&\frac{\rm d\sigma }{\rm
d\Omega}= \frac{ \lambcross^2}{32}\{\sin^2\theta[18|E1^1{\rm
P}_1|^2- 9|E1^3{\rm
P}_1|^2\nonumber\\
&&+9|M1^3{\rm D}_1|^2-25|E2^3{\rm D}_2|^2\nonumber\\
&&-18\sqrt2Re(M1^3{\rm S}_1^*\,M1^3{\rm D}_1)+ 30\sqrt3Re(M1^3{\rm
D}_1^*\,E2^3{\rm D}_2)\nonumber\\
&&+30\sqrt6Re(M1^3{\rm S}_1^*\,E2^3{\rm D}_2)\nonumber\\
&&+\cos\theta(60\sqrt3Re(E1^1{\rm P}_1^*\,E2^1{\rm D}_2)\nonumber\\
&&-60 Re(E1^3{\rm P}_1^*\,E2^3{\rm D}_2))\nonumber\\
&&+\cos^2\theta(150| E2^1{\rm D}_2|^2-100| E2^3{\rm D}_2|^2)]\nonumber\\
&&+\cos\theta[-12\sqrt6Re(E1^3{\rm P}_1^*\,M1^3{\rm S}_1)\nonumber\\
&&-12\sqrt3Re(E1^3{\rm P}_1^*\,M1^3{\rm D}_1)+60Re(E1^3{\rm
P}_1^*\,E2^3{\rm D}_2]\nonumber\\
&&+18| E1^3{\rm P}_1|^2+12| M1^3{\rm S}_1|^2+6| M1^3{\rm D}_1|^2\nonumber\\
&&+50| E2^3{\rm D}_2|^2+12\sqrt2Re(M1^3{\rm S}_1^*\,M1^3{\rm D}_1)\nonumber\\
&&-20\sqrt6Re(M1^3{\rm S}_1^*E2^3{\rm D}_2)-20\sqrt3Re(M1^3{\rm
D}_1^*\,E2^3{\rm D}_2)\}\,,\nonumber\\
\end{eqnarray}

where ${\lambda \hspace{-0.6em} {^-}}$  is the reduced wavelength of
the photon. (Notation: $^{2S+1}L_J$ , J is the total momentum of the
system).

The asymmetry $\Sigma(\theta_N)$ reaction cross-sections  is given
by the expression:

\begin{eqnarray}\label{eq2}
&&{\Sigma }({\theta })= \sin^2\theta\{18 \mid E1^1{\rm P}_1|^2-9|
E1^3{\rm P}_1|^2\nonumber\\
&&-9| M1^3{\rm D}_1|^2+25| E2^3{\rm D}_2|^2\nonumber\\
&&+18\sqrt2Re(M1^3{\rm S}_1^*\,M1^3{\rm D}_1)+ 10\sqrt3Re(M1^3{\rm
D}_1^*\,E2^3{\rm D}_2)\nonumber\\
&&+10\sqrt6Re(M1^3{\rm S}_1^*\,E2^3{\rm D}_2)\nonumber\\
&&+\cos\theta[60\sqrt3Re(E1^1{\rm P}_1^*\,E2^1{\rm D}_2)\nonumber\\
&&-60Re(E1^3{\rm P}_1^*\,E2^3{\rm D}_2)]\nonumber\\
&&+\cos^2\theta[150|E2^1{\rm D}_2|^2-100|E2^3{\rm
D}_2|^2]\}/\frac{32}{\lambcross^2}\frac{\rm d\sigma }{\rm d\Omega
}\,.
\end{eqnarray}

Expressions (1) and (2) can be represented in the following forms:

\begin{equation} \label {eq3} \frac{\rm d\sigma }{\rm d\Omega }=
A[{\sin^2\theta(1+\beta\cos\theta
+\gamma\cos^2\theta)+\varepsilon\cos\theta+\nu}]\,.\quad
\end{equation}

\begin{equation} \label {eq4}
 {\Sigma }({\theta }) = \frac
 {\sin^2\theta(1+\alpha+\beta\cos\theta+\gamma\cos^2\theta)}
 {\sin^2\theta(1+\beta\cos\theta+\gamma\cos^2\theta)+\varepsilon\cos\theta+\nu}.
\end{equation}

The coefficients $A$, $\alpha$, $\beta$, $\gamma$, $\varepsilon$,
and $\nu$ are expressed in terms of the multipole amplitudes as:

\begin{eqnarray}\label{eq5} && A = \lambcross^2/32\{18 | E1^1{\rm
P}_1|^2-9| E1^3{\rm P}_1|^2+9| M1^3{\rm D}_1|^2 \nonumber\\
&&-25| E2^3{\rm D}_2|^2 -18\sqrt2|M1^3{\rm S}_1||M1^3{\rm
D}_1|\cos[\delta(^3S_1)-\delta(^3D_1)]\nonumber\\
&&+30\sqrt6|M1^3{\rm S}_1||E2^3{\rm
D}_2|\cos[\delta(^3S_1)-\delta(^3D_2)]\nonumber\\
&&+30\sqrt3|M1^3{\rm D}_1||E2^3{\rm
D}_2|\cos[\delta(^3D_1)-\delta(^3D_2)]\};
\end{eqnarray}

where $\delta$  is the phase of the corresponding amplitude,

\begin{eqnarray}\label{eq6} &&\alpha =\{-18 | M1^3{\rm D}_1|^2
+50|E2^3{\rm D}_2|^2\nonumber\\
&&+36\sqrt2 | M1^3{\rm S}_1 || M1^3{\rm D}_1 | \cos[\delta(^3{\rm
S}_1) - \delta(^3{\rm D}_1)] \nonumber\\
&&-20\sqrt6 | M1^3{\rm S}_1 || E2^3{\rm D}_2 | \cos[\delta(^3{\rm
S}_1) - \delta(^3{\rm D}_2)]\nonumber\\
&&-20\sqrt3 | M1^3{\rm D}_1 || E2^3{\rm D}_2 | \cos[\delta(^3{\rm
D}_1) - \delta(^3{\rm D}_2)]\}\,/\,\frac{32}{\lambcross^2}A\,;
\end{eqnarray}

\begin{eqnarray}\label{eq7} &&\beta =\{60\sqrt3|E1^1{\rm
P}_1||E2^1{\rm
D}_2|\,cos[\delta(^1P_1)-\delta(^1D_2)]\nonumber\\
&&-60|E1^3{\rm P}_1||E2^3{\rm
D}_2|\,cos[\delta(^3P_1)-\delta(^3D_2)]\,/\,\}\frac{32}{\lambcross^2}A\,;
\end {eqnarray}

\begin {equation}\label{eq8} \gamma=\{150|E2^1{\rm
D}_2|^2-100|E2^3{\rm D}_2|^2\,/\,\frac{32}{\lambcross^2}A\,;
\end {equation}

\begin{eqnarray}\label{eq9} &&\varepsilon
=\{-12\sqrt3|E1^3{\rm P}_1||M1^3{\rm
D}_1|\,cos[\delta(^3P_1)-\delta(^3D_1)] \nonumber\\
&&-12\sqrt6|E1^3{\rm P}_1||M1^3{\rm
S}_1|\,cos[\delta(^3P_1)-\delta(^3S_1)]\nonumber\\
&&+60|E1^3{\rm P}_1||E2^3{\rm
D}_2|\,cos[\delta(^3P_1)-\delta(^3D_2)]\}\,/\,\frac{32}{\lambcross^2}A\,;\nonumber\\
\end{eqnarray}

\begin{eqnarray}\label{eq12} &&\nu =\{18 | E1^3{\rm
P}_1|^2+12| M1^3{\rm S}_1|^2 +6| M1^3{\rm D}_1|^2 \nonumber\\&&+50 |
E2^3{\rm D}_2|^2 +12\sqrt{2}|M1^3{\rm S}_1||M1^3{\rm
D}_1|\,cos[\delta(^3S_1)-\delta(^3D_1)]\nonumber\\
&&-20\sqrt{6}|M1^3{\rm S}_1||E2^3{\rm
D}_2|\,cos[\delta(^3S_1)-\delta(^3D_2)]\nonumber\\
&&-20\sqrt{3}|M1^3{\rm D}_1||E2^3{\rm
D}_2|\,cos[\delta(^3D_1)-\delta(^3D_2)]\}\,/\,\frac{32}{\lambcross^2}A.\nonumber\\
\end{eqnarray}

Thus, having determined the coefficients {\it A}, $\alpha$, $\beta$,
$\gamma$, $\varepsilon$, and $\nu$ from the measured data on the
differential cross section and the asymmetry of linearly polarized
photon reaction cross sections, one can gain information about the
contributions of individual multipole amplitudes to the reaction
cross section. The coefficient A represents the differential cross
section for the electric dipole transition with spin {\it S}=0 and
the contribution of spin {\it S}=1 transitions at the angle of
nucleon emission $\theta_N$=90$^0$. The coefficient $\gamma$ is
proportional to the contribution of the spin {\it S}=0 electric
quadrupole {\it E}2 transition. The coefficient $\beta$, describes
the interference between the electric dipole {\it E}1 and electric
quadrupole {\it E}2 amplitudes having spin {\it S}=0. The
coefficients $\alpha$, $\varepsilon$, and $\nu$ designate the
contributions from {\it S}=1 transitions of final-state particles.
Expression (3) leads to the following relations:
$\varepsilon$=[d$\sigma$(0$^0$)-d$\sigma$(180$^0$)]/2{\it A},
$\nu$=[d$\sigma$(0$^0$)+d$\sigma$(180$^0$)]/2{\it A}. Neglecting the
contribution of transitions with spin {\it S} = 1, we obtain that
$\varepsilon$=[d$\sigma$(0$^0$)-d$\sigma$
(180$^0$)]/2d$\sigma_1(90^0)$,
$\nu$=[d$\sigma$(0$^0$)+d$\sigma$(180$^0$)]/2d$\sigma_1(90^0)$ where
d$\sigma_1(90^0)$ is the differential cross section for the {\it
S}=0 electric  {\it E}1 transition at the angle of nucleon emission
$\theta_N$=90$^0$. Table\,1 shows distribution over the polar angle
of nucleon emission in c.m.s. for {\it E}1, {\it E}2 and {\it M}1
multipole transitions. It can be seen from Table\,I that for all the
mentioned transitions we have d$\sigma$(0$^0$)=d$\sigma$(180$^0$),
and hence should be about $\varepsilon$$\sim$0. If
$\varepsilon$$\not=$0, then this may be indicative of the
experimental errors, or of the insufficiency of {\it E}1, {\it E}2
and {\it M}1 approximation.

\vskip10pt

TABLE I: Polar-angle nucleon emission distribution for {\it E1},
{\it E2} and {\it M1} multipole transitions.
\begin{center}
\begin{tabular}[t]{c c c}
\hline\hline Spin of the & Multipole &    Angular \\ final-states &
transition & distribution  \\ \hline S=0  & $|E1^1{\rm P}_1|^2$ &
$\sin^2\theta$
\\ & $|E2^1{\rm D}_2|^2$ & $\sin^2\theta$$\cos^2\theta$\\ \hline &
$|E1^3{\rm P}_1|^2$ & 1+$\cos^2\theta$\\   S=1 & $|M1^3{\rm S}_1|^2$
& const \\  & $|M1^3{\rm D}_1|^2$ & 5-3$\cos^2\theta$\\ & $|E2^3{\rm
D}_2|^2$ &
1-3$\cos^2\theta$+4$\cos^4\theta$\\
\hline\hline
\end{tabular}
\end{center}

According to expression (4), the contributions of each of the spin
{\it S}=0 amplitudes lead to the asymmetry $\Sigma(\theta_N)$=1 at
all polar angles of nucleon emission, except the angles
$\theta_N$=0$^0$ and 180$^0$. The difference of the asymmetry
$\Sigma(\theta_N)$ from 1 may be due only to spin {\it S}=1
transitions. The smaller is the contribution from {\it S}=1
transitions, the more rectangular the reaction cross-section
asymmetry becomes. It is evident from expression (6) that if it is
the $E1^3P_1$ or $M1^3S_1$ amplitude is predominant, then we have
the coefficient $\alpha$ = 0, and according to the
$\Sigma(\theta_N)$ data for the reaction with linearly polarized
photons, the contributions of the mentioned two amplitudes are not
separated. One can see from expressions (6) and (10) that if the
$M1^3D_1$ amplitude is predominant, then

\begin {equation}\label{eq11} \alpha=-3\nu\,;
\end {equation}

and if the $E2^3D_2$ amplitude is dominant, then we have

\begin {equation}\label{eq12} \alpha=\nu\,;
\end {equation}

Thus, if the reaction cross section in collinear geometry is known
and knowing what particular {\it S}=1 transition is the basic one,
we can unambiguously calculate the asymmetry of the cross-section
$\Sigma(\theta_N)$ for the reaction with linearly polarized photons.

\section{ Experimental information review }

The measurement of reaction cross-section in the collinear geometry
presents a certain methodical problem. The first differential
cross-section measurements of the reactions \cite{4} led to the
conclusion about the absence of the isotropic component, i.e., about
the absence of transitions with parallel spins of final-state
particles. By now, the experimental evidence of the cross sections
has been obtained in both the $^4$He photodisintegration reactions
\cite{5,6,7,8,9} and the reactions of radiative capture of polarized
protons by tritium nuclei \cite{10,11}.

The differential cross sections for ($\gamma,p$) and ($\gamma,n$)
reactions in the 4$\pi$ geometry were measured \cite{1,5} with a
diffusion chamber placed in the magnetic field, at bremsstrahlung
photon energies ranging from the reaction threshold up to 150\,MeV.
Relying on those data, the LS-method was used in Ref.\,\cite{6} to
calculate the coefficients $A$, $\alpha$, $\beta$, $\gamma$,
$\varepsilon$, and $\nu$ (circles in Figs.\,1,2). It is apparent
from Fig.\,1 that in the photon energy range 20$\le
E_{\gamma}\le$100\,MeV the coefficient $\nu$ is independent of the
photon energy. The mean values of coefficients were found to be
$\overline\nu_p$ = 0.019$\pm$0.002, and $\overline\nu_n$
=0.028$\pm$0.003, and also $\overline\varepsilon_p$ = 0.$\pm$0.002
and $\overline\varepsilon_n$ =-0.001$\pm$0.003 (Fig.\,2) The errors
on the data points are statistical only. The obtained coefficients
$\varepsilon_p$$\sim$ $\varepsilon_n$$\sim$0 are in agreement with
the assumption that the {\it E}1, {\it E}2 and {\it M}1
approximation is sufficient for describing the available
experimental data, and the contribution from the higher multipoles
can be neglected.

\begin{figure}[h]
\noindent\centering{
\includegraphics[width=60mm]{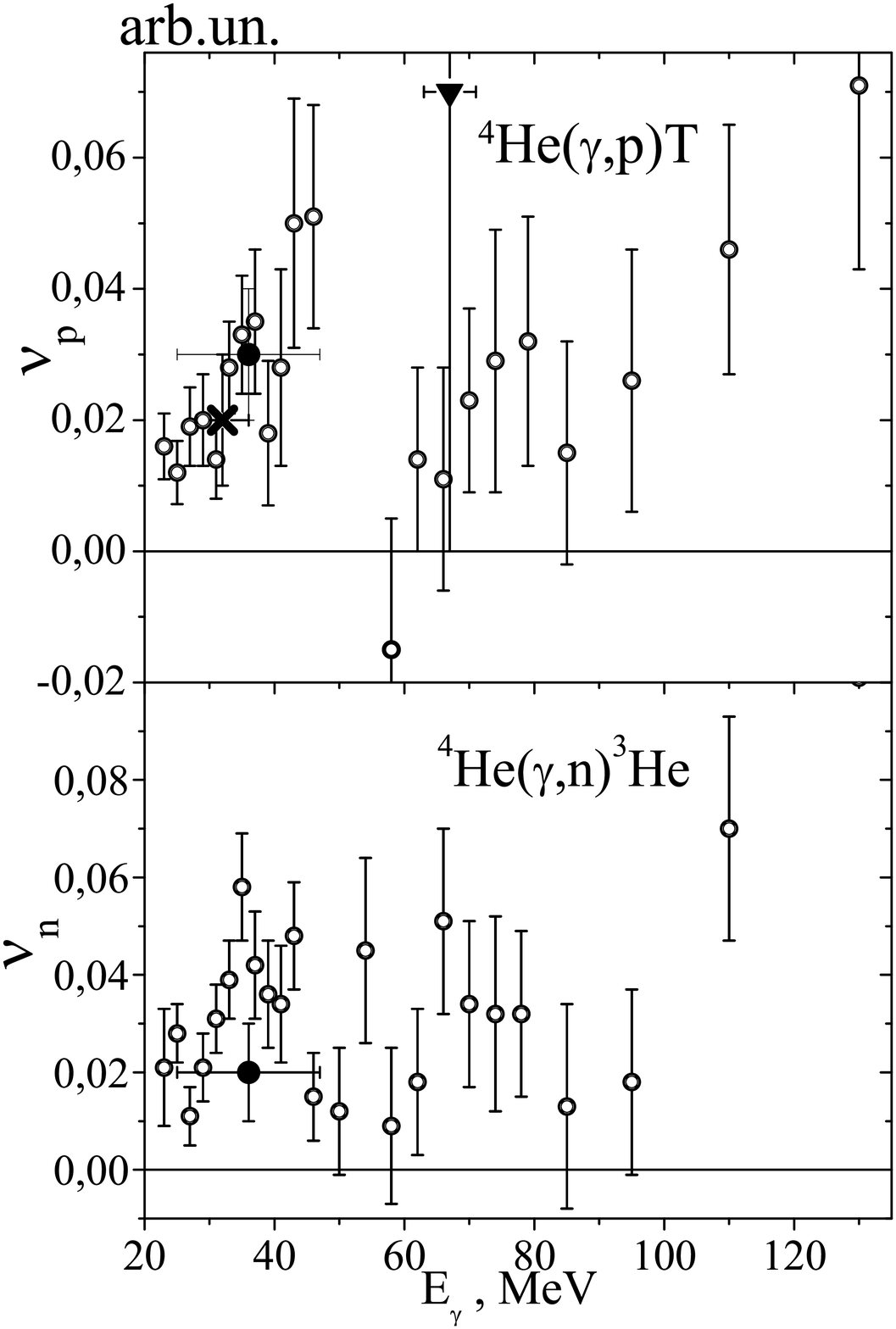}
} \caption{ The ratio of the total {\it S}=1 transition
cross-section in the collinear geometry to the {\it S}=0 electric
dipole transition cross-section at the nucleon emission angle
$\theta_N$=90$^0$. Open points show the data of Refs.\cite{1,5},
closed points - data of Ref. \cite{7}, the triangle - data of Ref.
\cite{8}, the cross - data from \cite{9}. The errors on the data
points are statistical only. }
\end{figure}

\begin{figure}[h]
\noindent\centering{
\includegraphics[width=60mm]{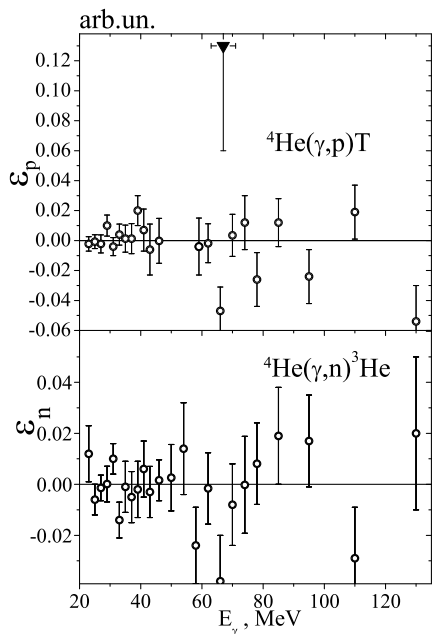}
} \caption{Coefficients  $\varepsilon_p$  and $\varepsilon_n$.
Points are the same as in Fig.1. Error bars are statistical only.}
\end{figure}

It was indicated in Ref.\,\cite{6} that the errors in the
measurements of the nucleon emission polar angles $\delta(\theta_N)$
result in the over-estimate of the coefficient $\nu$. In the limit,
when calculating the coefficient $\nu$ from the differential cross
section with d$\sigma$(0$^0$)=d$\sigma$(180$^0$)=0, i.e. $\nu$ = 0,
we obtain a coefficient $\nu>$0, which depends on the angular
resolution in the measurements of the nucleon emission polar angle.
By the use of simulation, the corresponding corrections were
calculated in Ref.\,\cite{6} on the assumption that the angular
resolution was $\delta(\theta_N)$=1$^0$ and the step of
histogramming the experimental data on the differential reaction
cross section was 10$^0$ \cite{5}. After taking into account those
corrections, the average values of the coefficients in the mentioned
photon energy range were determined to be
$\overline\nu_p$=0.01$\pm$0.002 and
$\overline\nu_n$=0.015$\pm$0.003. The difference between the
coefficients $\overline\nu_p$ and $\overline\nu_n$ may be due to the
fact that the errors in the measurement of the neutron emission
polar angle $\delta(\theta_n)$ were greater than the ones in
measuring the proton emission polar angle $\delta(\theta_p)$.

The closed points in Fig.1 show the data of Ref.\,\cite{7}. The
differential  cross-sections of the reaction were measured in the
4$\pi$ geometry on the bremsstrahlung beam at the maximum photon
energy E$_{\gamma}^{max}$=80\,MeV, using the diffusion chamber
placed in the magnetic field. The authors of Ref.\,\cite{7}
estimated the $\nu$ coefficients to be $\nu_p$=0.03$\pm$0.01 and
$\nu_n$ =0.02$\pm$0.01 in the photon energy region of the giant
dipole resonance. The triangle in Fig.\,1 shows the data of paper
\cite{8}. The measurements were made using tagged-bremsstrahlung
technique with photon of energy 67$\pm$4\,MeV. The ($\gamma,p$)
reaction was registered with the use of the large solid-angle
cylindrical detector based on a set of proportional wire chambers
and the scintillation-counter telescope in the range of proton
emission polar angles 35$^0$$\le\theta_p\le$140$^0$. Considering
that the detector did not register the reaction events at large and
small angles, the coefficients $\nu_p$ and $\varepsilon_p$ were
determined with large errors, $\nu_p$ =0.07$\pm$0.07 and
$\varepsilon_p$=0.13$\pm$0.07. The cross in Fig.\,1 shows the data
obtained in studies of $^4$He nuclear photodisintegration reactions
through the use of the 4$\pi$ time projection chamber in the photon
energy range 22.3$\le E_{\gamma}\le$32\,MeV \cite{9}. Based on the
measured differential $^4$He($\gamma,p$)T reaction cross section,
those authors have determined by the LS method that at
E$_\gamma$=32\,MeV the value makes $\nu_p$=0.02$\pm$0.01.

Important information about the {\it S}=1 transition cross-section
was derived from the reactions of radiative capture of polarized
protons by tritium nuclei. In \cite{10}, the reaction was
investigated at polarized proton energies $E_p$ between 0.86 and
9\,MeV.  The differential cross section and the analyzing power
$A_y$ were measured in the angular range
20$^0\le\theta_{\gamma}\le$155$^0$. It was inferred that the
$^3S_1M1$ transition was the basic {\it S}=1 transition. The average
ratio of this transition cross-section to the $^1P_1E1$ transition
cross-section was found to be $\nu$ = 0.006$\pm$0.004.

In investigation of the same reaction at the 2\,MeV proton energy
($E_{\gamma}$=21.25\,MeV), the differential cross section and the
analyzing power $A_y$ were measured in the angular range
0$^0\le\theta\le$155$^0$ \cite{11}. The multipole analysis has given
the cross sections for the $^1D_2E2$, $^3P_1E1$ and $^3D_2E2$
transitions in relation to the $^1P_1E1$ transition cross-section.
It is noted in the paper that the $^3S_1 M$1 transition can be the
basic one at the reaction threshold. However, this transition is
determined by the {\it S}-state of the particle system, the
contribution of which should decrease with energy increase as 1/V,
where V is the nucleon velocity. The conclusion was made about the
dominant contribution of the $^3P_1E1$ transition among spin {\it
S}=1 transitions. The $^3P_1E1$ transition cross-section has made up
0.72(+0.29-0.18)\% of the total cross section of the reaction.

It should be noted that in the studies of $^4$He nuclear
photodisintegration reactions \cite{7,8,9}, there were no
corrections made for the  angular resolution in the measurements of
nucleon emission polar angles, which could substantially reduce the
coefficient $\nu$ value. After taking these corrections into
account, the data on the coefficient $\nu$ obtained from the
reactions of photodisintegration of the $^4$He nucleus and from the
reaction of radiation capture of protons by tritium nuclei can be
agreed among themselves.

Summarizing the results of the above-considered works, the following
conclusions can be made. The true value of the coefficient $\nu$ may
lie within the limit $\nu$=0.010$\pm$0.005. The experimental data
show that in the photon energy range 20$\le E_{\gamma}\le$100\, MeV
the ratio of the {\it S}=1 transitions cross section in the
collinear geometry to the cross section of the {\it S}=0 electric
dipole transition at the nucleon emission angle $\theta_ N$=90$^0$
is independent of the photon energy, within the statistical error.

Figures 3 and 4 show the data on the angular/energy dependencies of
the asymmetry of cross-sections for the $^4$He($\vec\gamma,p$)T and
$^4$He($\vec\gamma,n$)$^3$He reactions with linearly polarized
photons. The polarization of the currently available photon beams is
estimated to be less than unity. Therefore, in experiments, the
product of the photon beam polarization $P_{\gamma}$ by the reaction
cross-section asymmetry $\Sigma(\theta_N)$ is measured, and
consequently, the asymmetry measurement errors include the
uncertainty in the measurement of photon beam polarization. The
circles in the Figures 3 and 4  show the results of work \cite{12}.
The linearly polarized photon beam was produced as a result of
coherent bremsstrahlung of electrons of energies E$_e$= 500, 600 and
800\,MeV in a diamond single crystal. The coherent bremsstrahlung
peaks were situated near the energies 40, 60 and 80\,MeV,
respectively. The degree of photon beam polarization was calculated
under the assumption of the unambiguous relationship between the
coherent effect value and the photon polarization \cite{13}. The
effective degrees of photon polarization in the energy ranges
34$<$E$_{\gamma}\le$46\,MeV, 46$<$E$_{\gamma}\le$65\,MeV and
65$<$E$_{\gamma}\le$90\,MeV were determined to be
$P_{\gamma}^{eff}$= 0.62, 0.71, and 0.75, respectively. The
statistical error being $\Delta$$P_{\gamma}^{eff}$=$\pm$0.03. The
events of $^4$He nuclear disintegration were registered using the
magnetic spectrometer with a helium streamer chamber.

The asymmetry of the $^4$He($\vec\gamma,p$)T reaction cross-section
was measured \cite{14} on the polarized photon beam, which resulted
from planar channeling of 1200\,MeV electrons in the diamond single
crystal (triangles in Fig.\,4). The calculations of the degree of
polarization were checked against the data, which were obtained when
measuring the asymmetry of deuteron photodisintegration
cross-section \cite{15}. The polarization decreased from 0.88 in the
range of the reaction threshold to 0.58 at E$_{\gamma}$=50\,MeV. The
reaction was registered with the use of the helium streamer chamber.
It has been concluded in \cite{14} that in the angular range
20$^0\le\theta_p\le$160$^0$ the asymmetry of the reaction cross
section is independent of the polar angle emission of the proton.

The squares in Fig.\,4 show the preliminary measurement data on the
asymmetry of the $^4$He($\vec\gamma,n$)$^3$He reaction cross section
as observed in Ref.\,\cite{14,16}. The experiment was done there at
energies 40$<$E$_{\gamma}\le$56\,MeV on the beam of polarized tagged
photons, which resulted from the coherent bremsstrahlung of
192.6\,MeV electrons in the diamond single crystal. The neutrons
were registered with a scintillation counter system using the
time-of-flight method. The measurements were carried out at the
neutron emission angles  $\theta_n$=45$^0$, 90$^0$ and 130$^0$.

\begin{figure}[h]
\noindent\centering{
\includegraphics[width=80mm]{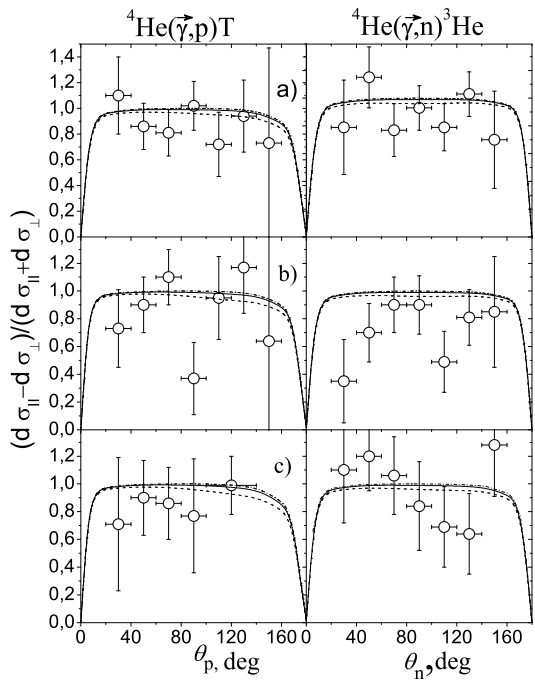}
} \caption{Angular dependences of the asymmetry of cross-sections
for the $^4$He($\vec\gamma$,p)T and $^4$He($\vec\gamma$,n)$^3$He
reactions with linearly polarized photons at energies of a) 40\,MeV,
b) 60\,MeV and c) 80\,MeV. Circles show the data of Ref.[12]. The
curves are explained in Fig.5.}
\end{figure}

\begin{figure}[h]
\noindent\centering{
\includegraphics[width=60mm]{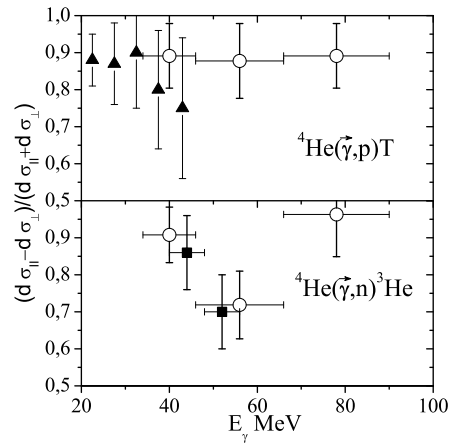}
} \caption{Energy dependences of the asymmetry of cross-sections for
the $^4$He($\vec\gamma$,p)T and $^4$He($\vec\gamma$,n)$^3$He
reactions with linearly polarized photons. Circles - data of. Ref.
[12], triangles - data of Ref.[14], squares - data of Ref.[14,16].}
\end{figure}

It is evident from Fig.\,4 that within the limits of experimental
error, the asymmetry $\Sigma(E_{\gamma}$) of the cross sections for
linearly polarized photon reactions is also independent of the
photon energy.
\begin{figure}[h]
\noindent\centering{
\includegraphics[width=80mm]{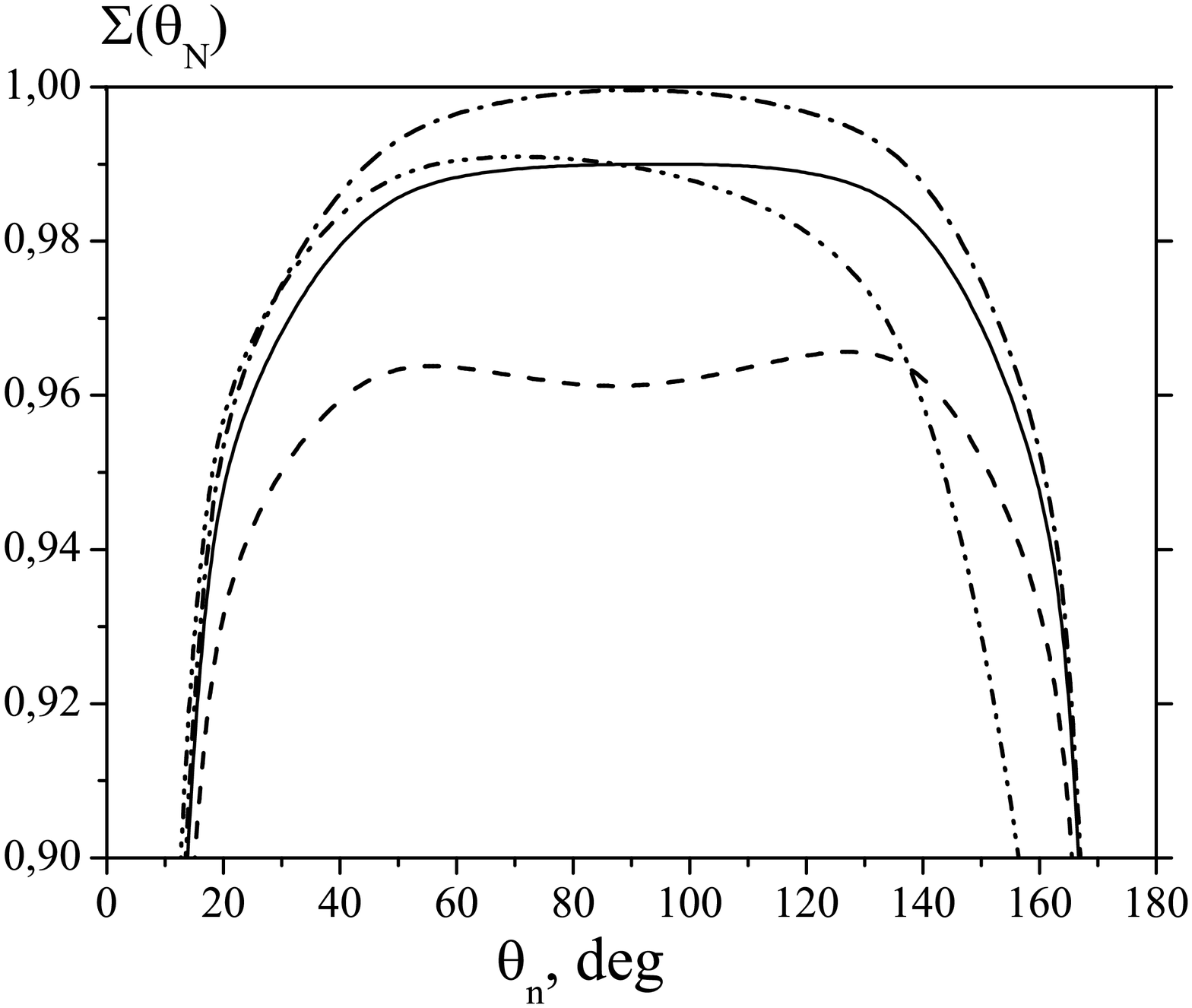}
} \caption{Angular dependencies for the $^4$He($\vec\gamma$,n)$^3$He
reaction cross-section asymmetry, calculated at photon energy of
40\,MeV and the coefficients $\varepsilon$=0 and $\nu$=0.01. The
solid curve was calculated with $E1^3P_1$ or $M1^3S_1$ assumed as
the basic transition; the dashed curve - with the basic $M1^3D_1$
transition; the dash-dot curve - with the basic $E2^3D_2$
transition; the dash-and-two dots curve -  with $E1^3P_1$ or
$M1^3S_1$ as the basic transition for the $^4$He($\vec\gamma$,p)T
reaction.}
\end{figure}

In computing the curves shown in Figs.\,3 and 5, the coefficients
$\beta$ and $\gamma$ were calculated by the LS method from the
differential cross sections for these reactions \cite{1,5}, and the
coefficients $\varepsilon$=0 and $\nu$=0.01 were used. Figure\,5
shows the calculation of the cross-section asymmetry
$\Sigma(\theta_n)$ of the $^4$He($\vec\gamma,n$)$^3$He reaction at
photon energy of 40\,MeV. The solid curve represents the calculation
under the assumption that $E1^3P_1$ or $M1^3S_1$ is the basic
transition; the dashed curve was calculated assuming that the basic
transition is $M1^3D_1$, the dash-and-dot curve - assuming $E2^3D_2$
to be the basic transition. The dash-and-two dots curve was
calculated on the assumption that $E1^3P_1$ or $M1^3S_1$ is the
basic transition for the $^4$He($\vec\gamma,p$)T reaction. The
reaction cross-section asymmetry $\Sigma(\theta_N$) is weakly
dependent on the coefficients $\beta$ and $\gamma$. For example, in
calculations of the solid curve the $\beta_n$=-0.05 and
$\gamma_n$=0.91 were used, and for the dash-and-two dots curve the
coefficients $\beta_p$=0.75 and $\gamma_p$=0.5 were used.

At the nucleon emission angle $\theta_N$=90$^0$ expression (4) leads
to simple relations. If $M1^3D_1$ is the basic transition, then we
have

\begin{equation} \label {eq13}
 {\Sigma }({90^0}) = \frac
 {1-3\nu}
 {1+\nu}.
\end{equation}
With the basic $E1^3P_1$ or $M1^3S_1$ transition, the asymmetry is
equal to:
\begin{equation} \label {eq14}
 {\Sigma}({90^0}) = \frac
 {1}
 {1+\nu}.
\end{equation}
and with the basic $E2^3D_2$ transition we have $\Sigma$(90$^0$)=1

It is obvious from Fig.\,3 that there is some disagreement between
the experimental data and the calculated curves for the
cross-section asymmetry $\Sigma(\theta_N)$. To determine the
discrepancy, the experimental data on the asymmetry
$\Sigma(\theta_N)$ \cite{12} were averaged in the interval of
nucleon emission polar angles 20$^0\le\theta_N\le$160$^0$ for the
both ($\vec\gamma,p$) and ($\vec\gamma,n$) reaction channels. The
asymmetry values calculated in the same interval of nucleon emission
polar angle were averaged for each of the possible {\it S}=1
transitions. Then the difference between the averaged asymmetry
values   $\overline
 {\Delta\Sigma}$=$\overline{\Sigma}_{th}$-$\overline{\Sigma}_{exp}$
was determined. The calculated results are given in Table\,II.

\vskip10pt

 TABLE II: Difference    between the averaged calculated/measured
asymmetries of cross sections for ($\vec\gamma$,p) and
($\vec\gamma$,n) reaction channels.
\begin{center}
\begin{tabular}[t]{cccc}
\hline\hline $E_{\gamma}$, & \multicolumn{3}{c} {$\overline
 {\Delta\Sigma}$=$\overline{\Sigma}_{th}$-$\overline{\Sigma}_{exp}$}
\\  MeV & $|M1^3{\rm D}_1|^2$ & $|E1^3{\rm P}_1|^2$ or$|M1^3{\rm S}_1|^2$&$|E2^3{\rm
D}_2|^2$\\
\hline
 40   & 0.055$\pm$0.067 &  $0.081\pm0.063$&$0.090\pm0.087$\\
60 & $0.166\pm0.068$ & $0.191\pm0.067$ & $0.201\pm0.067$\\  80 &
$0.017\pm0.086$ & $0.046\pm0.086$  & $0.056\pm0.086$\\

\hline\hline
\end{tabular}
\end{center}

Supposing that the cross-section asymmetry data are correct, then
the calculated difference $\overline\Delta\Sigma$ would lead to the
value $\nu>$0.04, which is not in agreement with the cross-section
data in the collinear geometry (see Fig.\,1). Besides, at this $\nu$
value, the angular dependence of the cross-section asymmetry would
manifest in an explicit form in the available statistics. The
discordance may be attributed to the instrumental errors in the
measurements of the reaction cross-section asymmetry. It should be
noted that in Refs. \cite{12}, \cite{14} and \cite{16} the reaction
products were registered by different methods, but, within the
experimental error, the results are in agreement. It is also
possible that there is the contribution of additional polarization
caused by the off-axis collimation of the photon beam. However, in
this case, the over-estimate of the degree of photon beam
polarization is possible only if the vectors of the polarizations
coincide, this being scarcely probable because the data were
obtained in several independent experiments. It is also possible
that the methods used to calculate the degree of polarization of
photon beams, as well as their verification by the
photodisintegration of the deuteron, can lead to an over-estimated
value of their degree of polarization.

The uncertainties in the available experimental data on the
asymmetry of $^4$He nuclear photodisintegration reaction
cross-sections give no way of drawing a conclusion about what
particular spin {\it S}=1 transition is dominant. Using the
conclusions of Refs. \cite{10} and \cite{11} that the basic
transition is, respectively, the $^3S_1M1$ transition and the
$^3P_1E1$ transition, which have the same asymmetry
$\Sigma(\theta_N)$, and also, the data on the reaction cross-section
in the collinear geometry $\nu$=0.01$\pm$0.005, we have calculated
the asymmetry of cross-sections for two-body reactions with linearly
polarized photons. It is found to be
$\Sigma(90^0)$=0.9901$\pm$0.005.

\section{ Discussion of results}

The available experimental data of two-body $^4$He nuclear
disintegration reactions allows us to suggest that the ratio of the
cross section of {\it S}=1 transitions in the collinear geometry to
the cross section of {\it E}1 {\it S}=0 transition at the nucleon
emission angle $\theta_N$ =90$^0$ in the photon energy range 20$\le
E_{\gamma}\le$100\,MeV is independent  of the photon energy, to
within the experimental errors. The cross-sectional asymmetry of the
reaction with linearly polarized photons in the energy range 20$\le
E_{\gamma}\le$90\,MeV is also independent of the photon energy, to
within the experimental errors.

In Ref.\,\cite{3}, the probabilities of the {\it S}, {\it P} and
{\it D} states of the $^4$He nucleus were computed. In the
computations, the authors used the realistic nucleon-nucleon CD-Bonn
and AV18 potentials, and also took into account the contribution of
3{\it N} forces in the form of the Tucson-Melbourne (TM) and Urbana
IX potentials. The computational results are given Table\,III.

\vskip10pt

TABLE III: {\it S}, {\it P} and {\it D} state probabilities for the
$^4$He wave functions. All probabilities are in percentage terms.
(The table is taken from Ref.\,[3])
\begin{center}
\begin{tabular}[t]{lccc}
\hline\hline Interaction & {\it S} & {\it P} & {\it D} \\
\hline
CD-Bonn   & 89.06 &  0.22 &10.72\\
CD-Bonn+TM & 89.65 & 0.45 & 9.90\\
AV18 & 85.87 & 0.35  & 13.78\\
AV18+Urb-IX & 83.23 & 0.75 &16.03\\
\hline\hline
\end{tabular}
\end{center}

It is evident from Table\,III that the consideration of 3{\it N}
forces nearly doubles the probability of {\it P} states. Therefore,
the measurement of {\it P}-state probability makes it possible to
obtain new information on 3{\it N} forces. It can be assumed that
when the nucleus is split using the electromagnetic interaction,
which is central, there is no spin flip of the hadron system of
particles. In the course of the reaction, contribution of the final
state interaction (FSI) is possible [17]. However, the spin-flip of
the particle system due to the FSI contribution is the second-order
quantity of smallness. Assuming also that the reaction cross section
does not depend on the spin state of the $^4$He nucleus, then it is
hoped that the ratio

\begin{equation} \label {eq15}
 {\eta} = \frac
 {^4He(^3P_0)\rightarrow \sigma(E1^3P_1)}
 {^4He(^1S_0)\rightarrow \sigma(E1^1P_1)}
\end{equation}

will be independent of the photon energy. Here $\sigma(E1^3P_1)$ is
the part of the ($\gamma,N$) reaction cross-section caused by the
spin {\it S}=1 transition, $\sigma(E1^1P_1)$ is the part of the
($\gamma,N$) reaction cross-section due to the electric dipole spin
{\it S}=0 transition. It can be assumed that

\begin{equation} \label {eq16}
 {\eta} \simeq {\mathscr{P}}(^3P_0)
\end{equation}
here ${\mathscr{P}}(^3P_0)$ is the probability of $^3P_0$ states of
the $^4$He nucleus.

Having integrated the angular distribution of the $E1^3P_1$
transition (see Table\,1) with respect to the solid angle, and
expressing its total cross section in terms of the differential
cross section at the nucleon emission angle $\theta_N$=0$^0$, we
obtained

 \begin{equation} \label {eq17}
 {\sigma(E1^3P_1)} = \frac
 {d\sigma(0^0)}
 {1+cos^2(0^0)}
 \frac{16\pi}{3}
\end{equation}

Similarly to the $E1^1P_1$ transition, we have
\begin{equation} \label {eq18}
 {\sigma(E1^1P_1)} = \frac
 {d\sigma(90^0)}
 {sin^2(90^0)}
 \frac{8\pi}{3}
\end{equation}
and the total cross-section ratio is
\begin{equation} \label {eq19}
 \frac
 { \sigma(E1^3P_1)}
 { \sigma(E1^1P_1)}=\frac{d\sigma(0^0)}{d\sigma_1(90^0)}=\nu
\end{equation}

If the $M1^3S_1$ transition is predominant, then $\eta$=3$\nu$/2,
and if the $M1^3D_1$ transition is predominant, then $\eta$=3$\nu$.

The calculation \cite{3} of the probabilities of $^3P_0$ states of
the $^4$He nucleus based on realistic {\it NN} potential, taking
into account the contribution of 3{\it N} forces and the available
experimental data on the cross section for transitions with spin
{\it S}=1, do not contradict each other.

In a number of theoretical \cite{2} and experimental \cite{10,11}
works the origin  of spin {\it S}=1 transitions is attributed to the
fact that in the course of the reaction the spin-flip of the
hadronic particle system takes place, and, in turn, the spin-flip is
caused by the contribution of MEC$^,$s. It should be noted that the
MEC$^,$s contribution is dependent on the photon energy \cite{18},
this being inconsistent with the given experimental data.

However, this does not preclude a MEC$^,$s contribution to spin {\it
S}=0 transitions of the final state of the particle system.

A systematic inconsistency is observed between the experimental data
on the reaction cross section in the collinear geometry and on the
cross-section asymmetry of the reaction with linearly polarized
photons. The inconsistency may be due, in particular, to the
overestimate of the calculated degree of photon beam polarization.
More information is needed to find out the source of the
inconsistency.

A high cross-section asymmetry value of ($\vec\gamma,N)$ reactions,
and also, its independence of the photon energy over a wide energy
range may make the $^4$He nucleus more convenient for measuring the
degree of linear photon-beam polarization than the deuteron will do
\cite{15}.

In this connection it should be noted that there is a need to
substantially improve the accuracy of angular dependence
measurements of the asymmetry of $^4$He photodisintegration reaction
cross-section, using  linearly polarized photons in the widest
possible range of photon energies.

The  author  wishes to  express his sincere gratitude to Dr. I.
Timchenko  for  useful  discussions  of this work.

\end{document}